\documentclass[twocolumn,showpacs,preprintnumbers,floatfix]{revtex4}
\usepackage{amssymb}
\usepackage{graphicx}
\usepackage{dcolumn}
\usepackage{bm}

\begin{document}

\title{Boson-Fermion pairing in a Boson-Fermion environment}
\author{A. Storozhenko,$^{1,2}$ P. Schuck,$^{1,3}$ T. Suzuki,$^4$ H. Yabu,$%
^4 $ J. Dukelsky$^5$}
\affiliation{$^1$Institut de Physique Nucl\'{e}aire, IN2P3-CNRS, Universit\'{e}
Paris-Sud, F-91406 Orsay C\'{e}dex, France \\
$^2$Bogoliubov Laboratory of Theoretical Physics, Joint Institute for
Nuclear Research, 141980 Dubna, Russia \\
$^3$Laboratoire de Physique et Mod\'elisation des Milieux Condens\'es, CNRS $%
\&$ Universit\'e Joseph Fourier, Maison des Magist\`eres, B.P. 166, 38042
Grenoble Cedex 9, France \\
$^4$Department of Physics, Tokyo Metropolitan University, 1-1 Minami-Ohsawa,
Hachioji, Tokyo 192-0397, Japan \\
$^5$Instituto de Estructura de la Materia, Serrano 123, Madrid 28006, Spain}
\date{\today}

\begin{abstract}
Propagation of a Boson-Fermion (B-F) pair in a B-F environment is
considered. The possibility of formation of stable strongly correlated
B-F pairs, embedded in the continuum, is pointed out.
The new Fermi gas of correlated B-F pairs shows a strongly modified Fermi
surface. The interaction between like particles is neglected in this
exploratory study. Various physical situations where our new pairing
mechanism could be of importance are invoked.
\end{abstract}

\maketitle



The physics of ultra cold atomic gases is making progress at a rapid pace,
which has led to a realization of Boson-Fermion mixtures of atomic gases
\cite{BF1,BF2,BF3,BF4}. Boson-Fermion mixtures may exhibit the richest
variety of phenomena of all. They may show very different behaviour from
pure Fermion or pure Bose gases \cite{BF-th1,BF-th2}. Especially interesting
is a possible instability of the mixture when there is an attraction between
Bosons and Fermions \cite{BF-th1,suz1,roth,capuz}, as the recent experiment
in fact suggests a collapse of the mixture \cite{coll}.

In the present work we propose and study quite a different scenario for an
attractively interacting Boson-Fermion mixture. To simplify the problem in a
first survey we shall consider the situation where there is no interaction
between atoms of the same kind. As we will discuss at the end of the paper,
this is not a severe approximation to cases where the interaction between like
atoms is repulsive.
More precisely we want to address the
question what happens to a mixture of free Fermions and Bosons when
a (tunable) attraction is switched on between Fermions and Bosons.
We imagine that correlated B-F pairs will be
created. These B-F pairs are composite Fermions and as such these B-F pairs
should form a new Fermi gas of composites. Besides in ultracold atomic gases
such a situation can exist in other branches of physics. For example in
nuclear systems (e.g. neutron stars) of high density $K^{-}$ - mesons and
nucleons may form a gas of $\Lambda $ 's and the $\Lambda $ 's may then form
a new Fermi gas of their own \cite{baldo}. Or in a quark gluon plasma
additional quarks may bind to preformed di-quarks or Color Cooper pairs (the
'bosons') \cite{aich} to form a new gas of nucleons in the so-called
hadronisation transition. Further examples may be added to this list.

For a numerical example, we take a mixture of $^{40}K$ - (fermion) and $^{41}K$
- (boson) atoms throughout the paper. They are known as candidates for a realization of such
kind of quantum systems. While their scattering lengths are not well fixed at present,
and different values have been reported experimentally \cite{potass}
it is not crucial at the moment because our study will be mostly
academic, elaborating on the basic phenomenon. Applications to realistic
systems shall be left for the future.

Let us consider a single B-F pair propagating in the background of a homogeneous
gas of free one component Fermions and spinless Bosons. We will formulate our
approach for a situation at finite temperature $T$, though later on in our
application we will concentrate on the $T=0$ case. We have in mind an analogous
study Cooper performed long time ago \cite{coop} for the propagation of two
fermions (spin up/down) in the background of a
a homogeneous gas of two component free Fermions. In other words we consider a
situation where in the original Cooper problem one Fermion type (let us say spin
down) is replaced by spinless bosons.
The B-F propagator at finite temperature $T$ and
finite centre of mass momentum $\mathbf{P}$ of the pair which is added to
the system with momenta $\mathbf{P}/2+\mathbf{p}$ (Fermion) and $\mathbf{P}%
/2-\mathbf{p}$ (Boson) is
\[
G_{\mathbf{p},\mathbf{p}^{\prime }}^{t-t^{\prime }}(\mathbf{P})=-i\theta
(t-t^{\prime })\langle \left\{ \left( b_{\frac{\mathbf{P}}{2}-\mathbf{p}}c_{%
\frac{\mathbf{P}}{2}+\mathbf{p}}\right) ^{t},\left( c_{\frac{\mathbf{P}}{2}+%
\mathbf{p^{\prime }}}^{+}b_{\frac{\mathbf{P}}{2}-\mathbf{p^{\prime }}%
}^{+}\right) ^{t^{\prime }}\right\} \rangle
\]%
where $\left\{ ,\right\} $ is the anticommutator and $c^{+}$ and $b^{+}$ are
Fermion and Boson creation operators, respectively. In ladder approximation
the integral equation for $G_{\mathbf{p},\mathbf{p}^{\prime }}(\mathbf{P},E)$
reads \cite{suz2}

\begin{eqnarray}
G_{\mathbf{p},\mathbf{p}^{\prime }}(\mathbf{P},E) &=&G_{\mathbf{p}%
}^{0}\left( \mathbf{P,}E\right) \delta (\mathbf{p}-\mathbf{p}^{\prime })
\nonumber \\
&&+\int \frac{d\mathbf{p}_{1}}{\left( 2\pi \right) ^{3}}G_{\mathbf{p}%
}^{0}\left( \mathbf{P,}E\right) V(\mathbf{p},\mathbf{p}_{1})G_{\mathbf{p}%
_{1},\mathbf{p}^{\prime }}(\mathbf{P},E)  \nonumber \\
&&
\end{eqnarray}%
In graphical form this equation is represented in Fig. 1.
\begin{figure}[tph]
\includegraphics[scale=0.5,angle=-90]{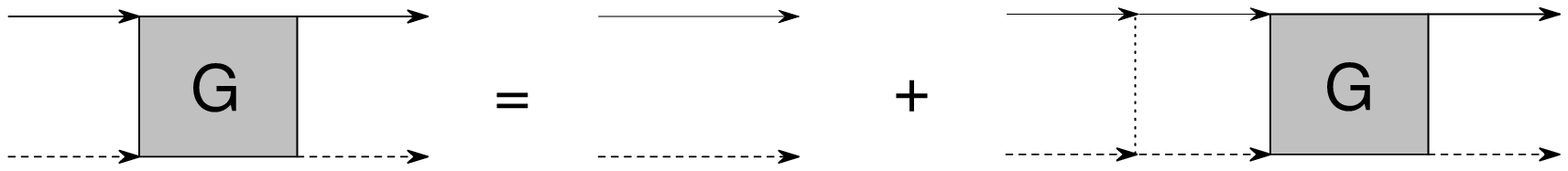}
\caption{Graphical representation of eq. (1). Dashed line stands for Boson,
straight line for Fermion. Dotted vertical line is interaction.}
\end{figure}

In (1) $V(\mathbf{p},\mathbf{p}_{1})$ is the B-F interaction and $G_{\mathbf{%
p}}^{0}\left( \mathbf{P,}E\right) $ is the free retarded B-F propagator in
the B-F background:
\begin{eqnarray}
G_{\mathbf{p}}^{0}\left( \mathbf{P,}E\right)  &=&\frac{1-f(\frac{\mathbf{P}}{%
2}+\mathbf{p})+g(\frac{\mathbf{P}}{2}-\mathbf{p})}{E-e_{f}\left( \frac{%
\mathbf{P}}{2}+\mathbf{p}\right) -e_{b}\left( \frac{\mathbf{P}}{2}-\mathbf{p}%
\right) +i\eta }  \nonumber \\
&&+\frac{\left( 2\pi \right) ^{3}n_{0}}{E-\frac{P^{2}}{2m}+i\eta }\delta (%
\frac{\mathbf{P}}{2}-\mathbf{p})
\end{eqnarray}%
Here $f(\mathbf{p})$ and $g(\mathbf{p})$ are the Fermi-Dirac and
Bose-Einstein distributions with chemical potentials $\mu _{f}$ and $\mu _{b}
$, respectively, and the term with the condensate fraction $n_{0}$ of Bosons
only appears for $T<T_{cr}$ where $T_{cr}$ is the critical temperature for
Bose condensation. We further have $e_{f}(p)=e_{b}(p)=p^{2}/2m$ which are
the kinetic energies of Fermions and Bosons which we suppose of equal mass: $%
m_{b}=m_{f}=m$. For simplicity we disregard mass shifts from selfenergy
corrections which may drive the masses of Fermions and Bosons apart, even if
in free space they are equal. Had we considered F-F propagation in a two component
Fermi gas (spin up/down), as Cooper did in his original work, than in (2)
the bosonic distribution $+g(\frac{\mathbf{P}}{2}-\mathbf{p})$ would have to be replaced by
$-f(\frac{\mathbf{P}}{2}-\mathbf{p})$ with, of course, $n_0 = 0$. As in Cooper's work,
equation (1) only treats the propagation of one pair and neglects the influence
of the other pairs on the pair under consideration. We therefore only can study situations
with a very low density of B-F pairs.

For the B-F case we will make the schematic
ansatz of separability of the force:
\begin{equation}
V(\mathbf{p},\mathbf{p}^{\prime })=-\lambda v(p)v(p^{\prime
}),\quad \lambda >0
\end{equation}%
with a Yukawa type of form factor
\[
v(p)=\frac{1}{\sqrt{m}(p^{2}+\beta ^{2})}
\]%
where, in principle, the two parameters $\lambda $ and $\beta $ may be related to the
scattering length and the effective range parameters of the low energy B-F
scattering in free space \cite{param}.
However, in this exploratory study
we will consider $\lambda $ and $\beta $ as free parameters especially in view of the fact
that the interaction strength can be shifted using the Feshbach resonance phenomenon,
whose application to K-atoms has been discussed in \cite{potass,bohn}.
The integral equation can then easily
be solved with only a quadrature to be done numerically. The result is
\begin{eqnarray}
G_{\mathbf{p},\mathbf{p}^{\prime }}(\mathbf{P},E) &=&G_{\mathbf{p}%
}^{0}\left( \mathbf{P,}E\right) \delta (\mathbf{p}-\mathbf{p}^{\prime })
\nonumber \\
&&-\frac{1}{\left( 2\pi \right) ^{3}}\frac{\lambda G_{\mathbf{p}}^{0}\left(
\mathbf{P,}E\right) v(p)v(p^{\prime })G_{\mathbf{p}^{\prime }}^{0}\left(
\mathbf{P,}E\right) }{1+\lambda J_{0}(E,\mathbf{P})}  \nonumber \\
&&
\end{eqnarray}%
where
\begin{equation}
J_{0}(E,\mathbf{P})=\int \frac{d\mathbf{p}}{\left( 2\pi \right) ^{3}}G_{%
\mathbf{p}}^{0}\left( \mathbf{P,}E\right) v^{2}(p)
\end{equation}%
Without loss of generality we can consider the simpler propagator integrated
over relative momentum
\begin{eqnarray}
G(\mathbf{P},E) &=&\int \frac{d\mathbf{p}^{\prime }}{\left( 2\pi \right) ^{3}%
}\int d\mathbf{p}v(p)v(p^{\prime })G_{\mathbf{p},\mathbf{p}^{\prime }}(%
\mathbf{P},E)  \nonumber \\
&=&\frac{J_{0}(E,\mathbf{P})}{1+\lambda J_{0}(E,\mathbf{P})}
\end{eqnarray}%
We will be interested in the $T$-matrix \cite{fett}
\begin{equation}
T_{E}^{\mathbf{P}}\left( \mathbf{q},\mathbf{q}^{\prime }\right) =\frac{%
-\lambda v(q)v(q^{\prime })}{1+\lambda J_{0}(E,\mathbf{P})}
\end{equation}%
and want to study the pole structure of this function, first at $T=0$, as a
\begin{figure}[tph]
\includegraphics[scale=0.3,angle=-90]{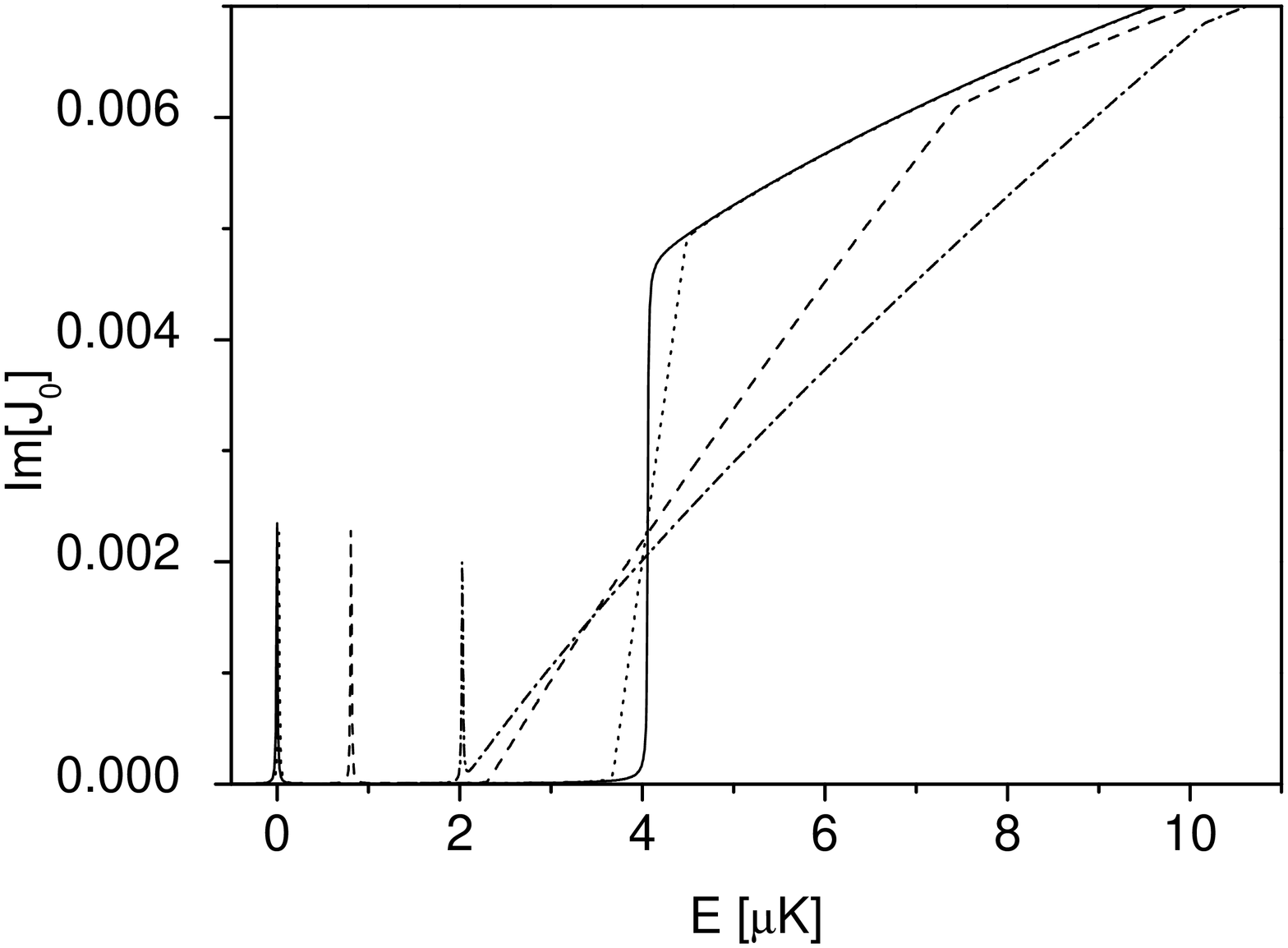}
\caption{$\mathop{\rm Im}J_{0}$ as a function of energy $E$ for different
values of the total momenta. Solid line corresponds to the $P^2/2m = 0$;
dotted line - $P^2/2m = 0.01\mu_f$; dashed line - $P^2/2m = 0.4\mu_f$
and dash-dotted line - $P^2/2m = \mu_f$.}
\end{figure}
function of $\mathbf{P}$. To this purpose we show in Fig.2 the imaginary
part of $J_{0}$\ as a function of $E$ for different values of the center of
mass momentum $\mathbf{P}$.
Calculation have been done for a $^{40}K - ^{41}K$ system with equal masses:
$m=m_{B}=m_{F}=0.649\times 10^{-22}g$.
We choose a situation where
the number of bosons is much less than the one of fermions:
$n_{F}=10^{14}cm^{-3},n_{B}=0.004n_{F}$. In this way
the B-F pair density will be very low and the
perturbation of the bosons on the fermions can be kept relatively small,
justifying the single pair ansatz (1).
For the B-F forces (3) we used $\beta =-7.89\frac{10^{-3}}{a_{B}}$ and
$\lambda $ =$\lambda ^{\prime }\frac{10^{-9}}{(2\pi a_{B})^{3}}$ where $a_{B}
$ is the Bohr radius and $\lambda ^{\prime }$ is the intensity of the force
which shall be varied in certain limits.

We see that for $P=0$, i.e. the B-F
pair being at rest, the imaginary part is zero below $2\mu _{f}=4.06\mu K$.
Exactly at $E=2\mu _{f}$\ the imaginary part jumps to a finite value,
increasing afterwards with the free gas law. For finite $\mathbf{P}$ the
imaginary part \ invades the region below $2\mu _{f}$\ and \ starts with a
finite slope. The threshold $E_{thr}$ follows the law
\[
2mE_{thr}=2k_{F}^{2}+P^{2}-2Pk_{F}
\]%
with $k_{F}=\sqrt{\frac{2m}{\hbar ^{2}}\mu _{f}}$ the Fermi momentum. We
also see a sharp peak at $E=P^{2}/2m$\ (the finite width is numerical). This
peak corresponds to the motion of the free Fermions when the Bosons are at
rest in the condensate. We can call this peak the one of the free B-F
pairs. The corresponding free B-F propagator is
\begin{eqnarray}
G_{n_{0}}^{0}(P,E) &=&\frac{n_{0}}{E-P^{2}/2m+i\eta }  \nonumber \\
&=&\mathcal{P}\frac{n_{0}}{E-P^{2}/2m}-i\pi n_{0}\delta \left(
E-P^{2}/2m\right)   \nonumber \\
&&
\end{eqnarray}%
We see that this part of the propagator is equal to the pure single Fermion
propagator multiplied with $n_{0}$ which is the free boson propagator at $T=0
$. The peak of (8) and the threshold of the continuum part of $%
\mathop{\rm
Im}J_{0}$ approach one another for increasing $\mathbf{P}$ and meet exactly
at $E=k_{F}^{2}/2m$ when the free BF pair moves with $P=k_{F}$.

\begin{figure}[tph]
\includegraphics[scale=0.3,angle=-90]{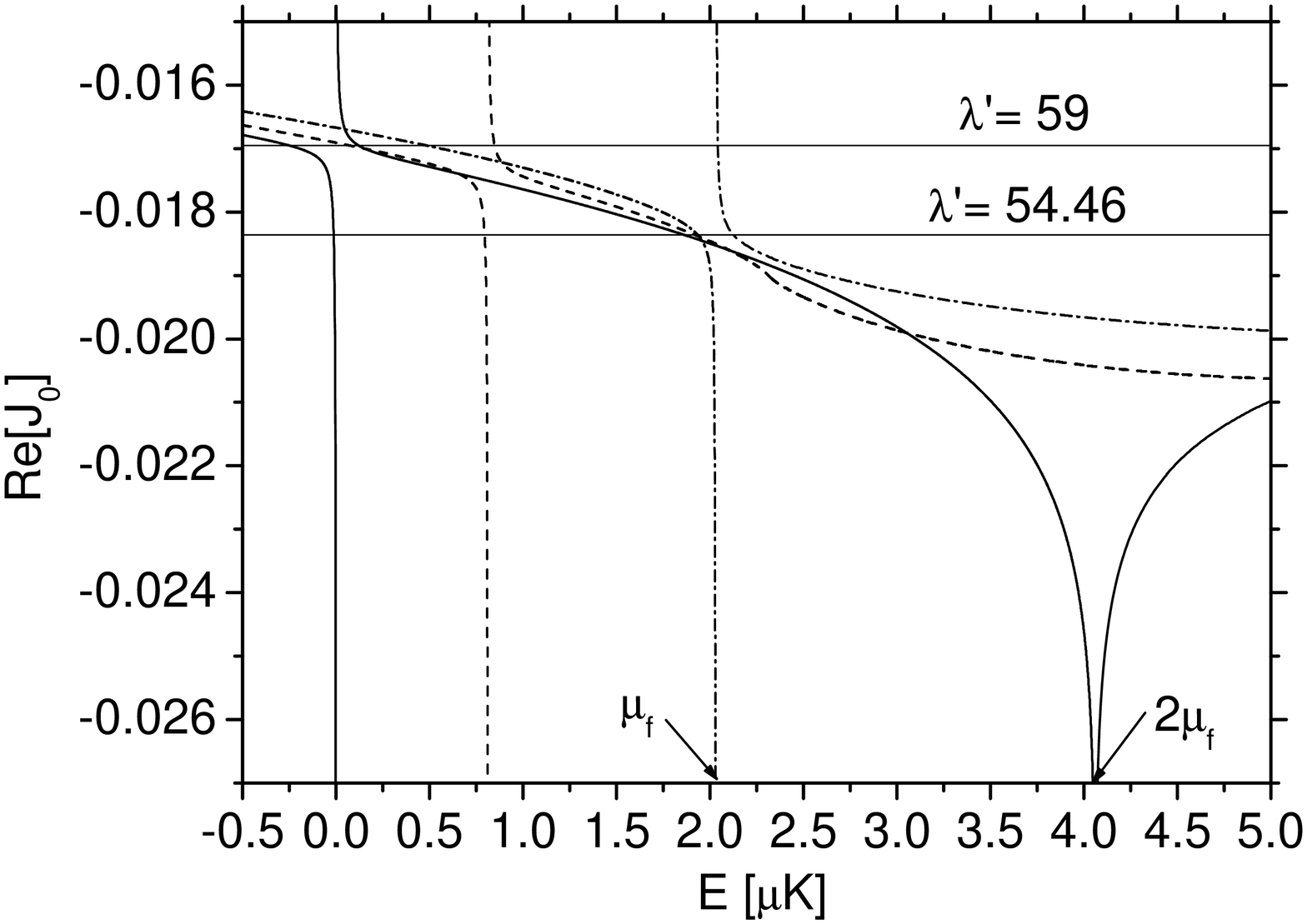}
\caption{$\mathop{\rm Re}J_{0}$ as a function of energy $E$ for different
values of the total momenta. Solid line corresponds to the $P^2/2m = 0$;
dashed line - $P^2/2m = 0.4\mu_f$ and dash-dotted line - $P^2/2m = \mu_f$.
Intersection with horizontal lines $-1/\protect%
\lambda ^{\prime }$ are also shown.}
\end{figure}
In Fig.3 we show the real part of $J_{0}$\ at $T=0$ for different values of $%
\mathbf{P}$. The poles of the $T$ - matrix (7) are determined by $%
\mathop{\rm Re}J_{0}=-\lambda ^{-1}$, i.e. by the intersection of $%
\mathop{\rm Re}J_{0}$ with the horizontal line $-\lambda ^{-1}$. We can see
that attractive potentials always lead to two stable solutions for each $%
\mathbf{P}$, as long as $\lambda $ does not become too large or too small.
Each of these solutions is defined by contributions of two parts of $J_{0}.$
One of them which is given by the integral part of $J_{0}$ we call the
collective contribution, the other one which comes from the $\frac{%
n_{0}}{E-P^{2}/2m}$ part of $J_{0}$ we call the ordinary or free one. As
long as both roots are well separated in energy, they can be determined by
the two separate dispersion equations
\begin{equation}
\mathcal{P}\int \frac{d\mathbf{p}}{\left( 2\pi \right) ^{3}}\frac{1-f(\frac{%
\mathbf{P}}{2}+\mathbf{p})+g(\frac{\mathbf{P}}{2}-\mathbf{p})}{E-e_{f}\left(
\frac{\mathbf{P}}{2}+\mathbf{p}\right) -e_{b}\left( \frac{\mathbf{P}}{2}-%
\mathbf{p}\right) }v^{2}(p)=-\frac{1}{\lambda }
\end{equation}%
\begin{equation}
\frac{1}{E-P^{2}/2m}=-\frac{1}{\lambda n_{0} v^{2}(P/2)}
\end{equation}
In this work we will restrict to $T=0$ when the phase space factors in (9)
reduce to $g(\textbf{p})=0$ and $f(\textbf{p})=\theta(k_f^2-p^2)$.
\begin{figure}[tph]
\includegraphics[scale=0.3,angle=-90]{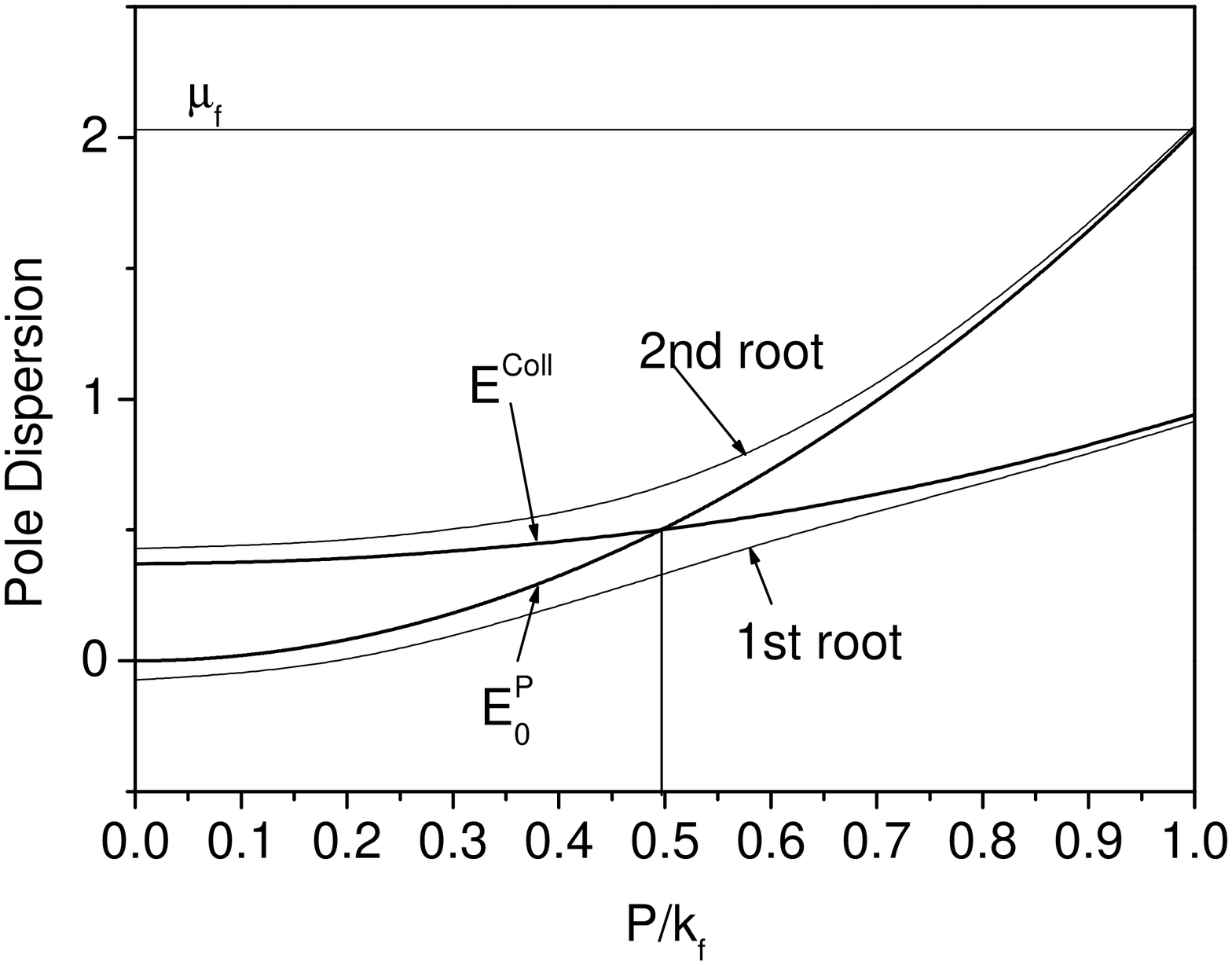}
\caption{Dispersion of the two true poles (thin lines) at $\protect\lambda %
^{\prime }=58$ in comparision with dispersion of pure collective pole $E^{Coll}$
and the free B-F pair $E_{0}^{P}$ (thick lines).}
\end{figure}

In Fig. 4 we show for $\lambda ^{\prime }=58$ the dispersion of the
collective pole (9) and of the ordinary pole (10) which describes to very good
approximation (with the parameters used here $\lambda n_{0} v^{2}(P/2) \ll 1$) the
center of mass motion of a non interacting B-F pair i.e. $E_{0}^{P}=P^{2}/2m$
(thick lines). On the same figure we also show the true dispersion of the
two roots (thin lines). We notice that at $P\simeq k_{F}/2$ there is a level
crossing between (9) and (10). However, in reality, due to the no crossing
rule \cite{cross1} and the level-level\ repulsion the two roots do \textit{%
not} cross but, as well known \cite{cross1}, nevertheless exchange their
character around crossing. For $P\lesssim k_{F}/2$ the collective pole is above
the ordinary root whereas for $P\gtrsim k_{F}/2$ it is the inverse. This
interchange has dramatic consequences: all B-F pairs with center of mass
momenta $k_{F}/2\lesssim P<k_{F}$ will populate the lower branch, i.e. the
collective pole. Due to its strong collectivity the upper part of the Fermi
sphere becomes strongly modified, as we will see later.
Of course, this interpretation is qualitative, since we
only considered a single B-F pair and, as in the case of ordinary Cooper
pairing \cite{schr}, pair-pair interaction may modify the scenario
quantitatively. How much of the original free Fermi surface melts and turns
\begin{figure}[tph]
\includegraphics[scale=0.3,angle=-90]{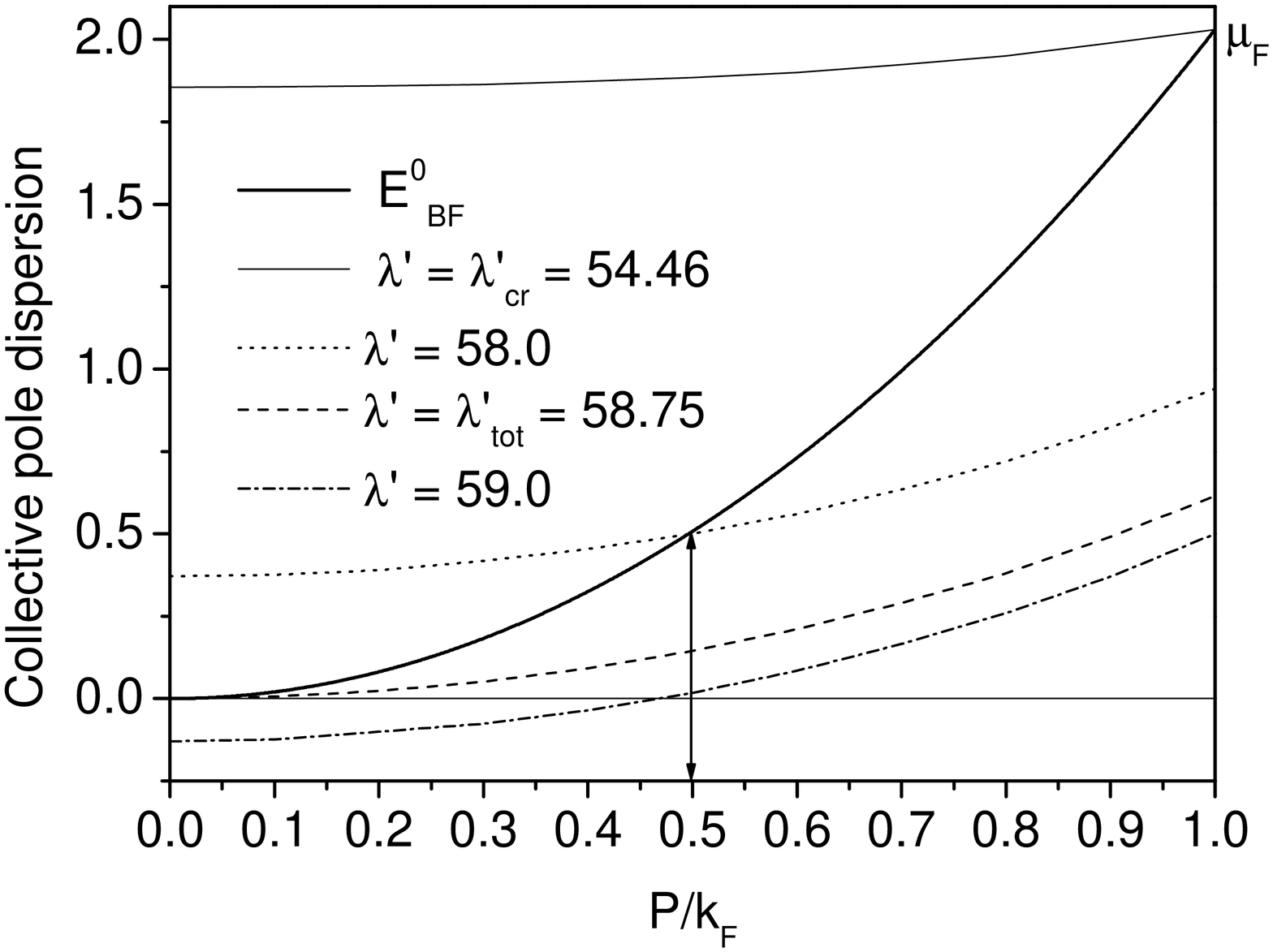}
\caption{Dispersion of the collective poles at different values of interaction
strength, together with the free B-F dispersion $E_{0}^{P}$.}
\end{figure}
into a new momentum space shell filled with a gas of B-F pairs
depends, of course, on the interaction. For $\lambda ^{\prime }<58$ the
new shell will be thinner than the one of Fig. 4 and at $\lambda
^{\prime }=\lambda _{cr}^{\prime }$ the shell of B-F pairs disappears. For
the parameters used here this happens for $\lambda _{cr}^{\prime }=54.46$.
One may, however, also define another critical value $\lambda ^{\prime
}=\lambda _{tot}^{\prime }$ which corresponds to the conversion of
practically all original bosons into B-F pairs. For our case this
occurs at $\lambda _{tot}^{\prime }\simeq 58.75$. Increasing the interaction
further, part of the B-F pairs will be converted into bound B-F molecules
with negative binding energy. The various scenarios are depicted in Fig. 5
where we show the dispersion of the collective pole (9) in comparison with the
free gas B-F dispersion $E_{0}^{P}=P^{2}/2m$ for four cases $\lambda
^{\prime }=\lambda _{cr}^{\prime }=54.46,$ $\lambda ^{\prime }=58,$ $\lambda
^{\prime }=\lambda _{tot}^{\prime }\simeq 58.75$ and $\lambda ^{\prime }=59$%
. We see that for $\lambda ^{\prime }=\lambda _{tot}^{\prime }$ the new
dispersion of the B-F pairs undershoots the free gas B-F dispersion
everywhere and that for $\lambda ^{\prime }=59$
molecules appear in the range $0\leq P\lesssim 0.47k_{F}$.

For $\lambda ^{\prime }<\lambda _{cr}^{\prime }$ still a \textit{stable} pole
i.e with no imaginary part exists
down to infinitesimally small attraction where for $P=0$ the collective pole
hits the value $2\mu _{f}$. This is due to the logarithmic divergency seen
in Fig. 3, which is of the same origin as in the original Cooper problem of
Fermions \cite{coop}, namely the sharp Fermi function in (9) at $P=T=0$. In
regions where the collective pole lies above the ordinary B-F pole one would
call the collective pole a B-F pair vibration which for $\lambda ^{\prime
}\lesssim \lambda _{cr}^{\prime }$ can become of considerable collectivity as
seen in Fig. 5. Whether like in nuclear physics \cite{cross1} such pair
vibrations can be detected experimentally is an open question.

To evaluate the ratio of the fermions and bosons which participate in the
composite BF pair let us consider the fermion (boson) occupation numbers%
\begin{equation}
n_{\mathbf{p}}\left( \mu ,T\right) =\int \frac{dE}{2\pi }f\left( E\right) A(%
\mathbf{p},E)
\end{equation}%
which can be found by the help of the single particle spectral function $%
A\left( \mathbf{p},E\right) $%
\begin{equation}
A_{b,f}\left( \mathbf{p},E\right) =\frac{-2\mathop{\rm Im}\Sigma
_{b,f}^{ret}(E,\mathbf{p})}{\left( E-e_{\mathbf{p}}^{b,f}-\mathop{\rm Re}%
\Sigma _{b,f}^{ret}(E,\mathbf{p})\right) ^{2}+\mathop{\rm Im}\Sigma
_{b,f}^{ret}(E,\mathbf{p})^{2}}
\end{equation}%
calculated via the retarded self-energy $\Sigma _{b,f}^{ret}\left( E,\mathbf{%
p}\right) $. The retarded self-energy can be defined \ through the Matsubara
self energies $\Sigma _{b,f}(z_{n},\mathbf{p})$
(again we keep the formalism general and work at finite temperature; however,
at the end we set $T=0$).
To find $\Sigma
_{b,f}(z_{n},\mathbf{p})$ we express them in terms of the $T$ matrix
calculated in ladder approximation
\begin{equation}
\Sigma _{b,f}(z_{n},\mathbf{p})=\pm T\sum_{z_{n^{\prime }}}\int \frac{d%
\mathbf{p}^{\prime }}{\left( 2\pi \right) ^{3}}T_{z_{n}+z_{n^{\prime }}}^{%
\mathbf{K}}\left( \mathbf{q},\mathbf{q}\right) G_{f,b}^{0}\left(
z_{n^{\prime }},\mathbf{p}^{\prime }\right)
\end{equation}%
\bigskip where $\mathbf{K}=\mathbf{p}+\mathbf{p}^{\prime }$, $\mathbf{q}=(%
\mathbf{p}-\mathbf{p}^{\prime })/2$ and $G_{b,f}^{0}\left( z_{n^{\prime }},%
\mathbf{p}^{\prime }\right) $ are the free single particle Green's functions%
\begin{eqnarray}
G_{b}^{0}\left( z_{n},\mathbf{p}\right) &=&\frac{1}{iz_{n}-e_{\mathbf{p}}^{b}%
}-\left( 2\pi \right) ^{3}T^{-1}n_{0}\delta \left( \mathbf{p}\right) \delta
_{n,0} \\
G_{f}^{0}\left( z_{n},\mathbf{p}\right) &=&\frac{1}{iz_{n}-e_{\mathbf{p}}^{f}%
}  \nonumber
\end{eqnarray}

The sum over the Matsubara frequencies can be performed using the spectral
representation of the T-matrix and the s.p. GF and transforming the sum into
contour integral . The corresponding imaginary parts will be%
\begin{eqnarray}
\mathop{\rm Im}\Sigma _{b}^{ret}(E,\mathbf{p}) &=&\int \frac{d\mathbf{p}%
^{\prime }}{\left( 2\pi \right) ^{3}}\mathop{\rm Im}T_{e_{\mathbf{p}^{\prime
}}^{f}+E}^{\mathbf{K}}\left( \mathbf{q},\mathbf{q}\right) \times  \nonumber
\\
&&\left[ f\left( e_{\mathbf{p}^{\prime }}^{f}\right) -f\left( e_{\mathbf{p}%
^{\prime }}^{f}+E\right) \right]
\end{eqnarray}

\begin{eqnarray}
\mathop{\rm Im}\Sigma _{f}^{ret}(E,\mathbf{p}) &=&n_{0}\mathop{\rm Im}%
T_{E+i0}^{\mathbf{p}}\left( \mathbf{p}/2,\mathbf{p}/2\right)  \nonumber \\
&&+\int \frac{d\mathbf{p}^{\prime }}{\left( 2\pi \right) ^{3}}\mathop{\rm Im}%
T_{e_{\mathbf{p}^{\prime }}^{b}+E+i0}^{\mathbf{K}}\left( \mathbf{q},\mathbf{q%
}\right) \times  \nonumber \\
&&\left[ g\left( e_{\mathbf{p}^{\prime }}^{b}\right) +f\left( e_{\mathbf{p}%
^{\prime }}^{b}+E\right) \right]
\end{eqnarray}%
The real parts can be calculated from the imaginary ones by using the
dispersion relation%
\begin{equation}
\mathop{\rm Re}\Sigma (E,\mathbf{p})=\Sigma ^{0}\left( \mathbf{p}\right) +%
\mathcal{P}\int \frac{dE^{\prime }}{\pi }\frac{\Gamma (E^{\prime },\mathbf{p}%
)}{E-E^{\prime }}
\end{equation}%
where $\Gamma (E,\mathbf{p})=i\left( \Sigma ^{ret}(E,\mathbf{p})-\Sigma
^{adv}(E,\mathbf{p})\right) $.

For the energies below $E^{thr}$ the quantity $J_{0}^{I}(E,K)$ is zero and
the $T$- matrix exhibits poles at $1+\lambda J_{0}^{R}(E,K)=0.$ The
corresponding value of $\mathop{\rm Im}T_{e_{\mathbf{p}^{\prime
}}^{b}+E+i0}^{\mathbf{K}}\left( \mathbf{q},\mathbf{q}\right) $ in this case
is the following
\begin{equation}
\mathop{\rm Im}T_{E+i0}^{\mathbf{K}}\left( \mathbf{q},\mathbf{q}\right)
=V\left( \mathbf{q},\mathbf{q}\right) \sum_{\nu }A_{\nu }^{K}\delta \left(
E-E_{\nu }^{K}\right)
\end{equation}%
where
\begin{equation}
A_{\nu }^{K}=\left[ \lambda \frac{dJ_{0}^{R}(E,K)}{dE}\right] _{E=E_{\nu
}^{K}}^{-1}
\end{equation}%
and $E_{\nu }^{K}$ is the solution of the secular equation $1+\lambda
J_{0}^{R}(E_{\nu }^{K},K)=0.$

For the calculation of the occupation numbers we use the quasiparticle
approximation (we checked this to be very accurate)

\begin{equation}
A_{b,f}\left( \mathbf{p},E\right) =\frac{2\pi \delta \left( E-\xi
_{p}^{b,f}\right) }{\left\vert 1-\frac{d}{dE}\mathop{\rm Re}\Sigma
_{b,f}^{ret}(E,\mathbf{p})\right\vert _{E=\xi _{p}^{b,f}}}
\end{equation}%
\bigskip where $\xi _{p}^{b,f}$ is the solution of the following equation%
\begin{equation}
\xi _{p}^{b,f}-e_{\mathbf{p}}^{b,f}-\mathop{\rm Re}\Sigma _{b,f}^{ret}(\xi
_{p}^{b,f},\mathbf{p})=0
\end{equation}

\begin{figure}[tph]
\includegraphics[scale=0.5,angle=0]{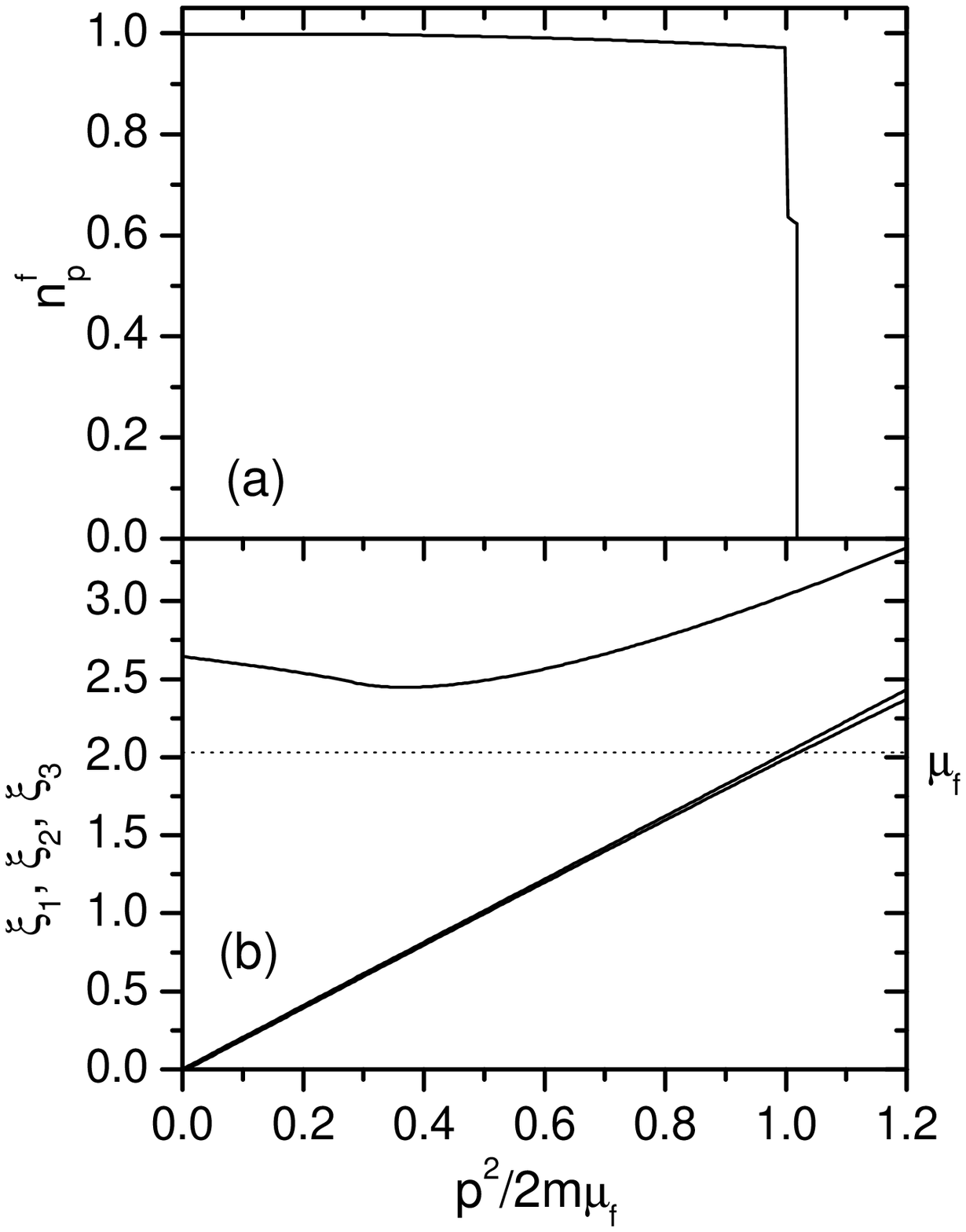}
\caption{Fermion occupation numbers (a) and dispersion of the quasifermion
energies (b) at $\protect\lambda ^{\prime }=52.0$.}
\end{figure}
The fermionic distribution function $n_{p}^{f}\left( \mu _{f},T\right) $ was
calculated at zero temperature in the approximation that $\mathop{\rm Im}%
\Sigma _{f}^{ret}(E,\mathbf{p})$ of eq. (16) contains only the term with the
boson condensate. We estimated that this term gives the by far largest
contribution in comparison with the second term. At weak interaction the
redistribution of the fermions due to their interaction with bosons is
small. As an example, in Fig.6a we show $n_p^f$ calculated
for $\lambda ^{\prime }= 52.0$.
We can see that the usual Fermi step function of the free fermions converts
into two steps. To
interpret such a behaviour we show in Fig. 6b the dispersion of the
quasifermion energies, solutions of (21). Since the mass operator (16)
contains via the $T$ - matrix the two poles we have discussed in Figs. 3, 4,
eq. (21) will have three roots which in the weak coupling limit considered
here correspond approximately to the free solution $\xi _{1p}\sim e_{p}^{f}$,
a second one $\xi _{2p}$ corresponding to (10) and almost degenerate with $%
\xi _{1p}$ up to the level crossing, and a third one which corresponds to the
collective BF pair (9) corresponding to the highest root in Fig. 6b.
Since our interaction strength is still
sufficiently weak so that the original chemical potential is barely changed,
the intersection at $p^{2}/2m\mu _{F}\simeq 1.02$ of the $\xi _{2p}$ with $%
\mu _{F}$ gives the upper step in Fig. 6a whereas the solution $\xi
_{1p}$ still intersects at $p^{2}/2m\simeq \mu _{F}$ and gives raise to the
intermediate step in Fig. 6a.
When the interaction further decreases, the two lowest roots
become completely degenerate and we have the usual Fermi step.

\begin{figure}[tph]
\includegraphics[scale=0.5,angle=0]{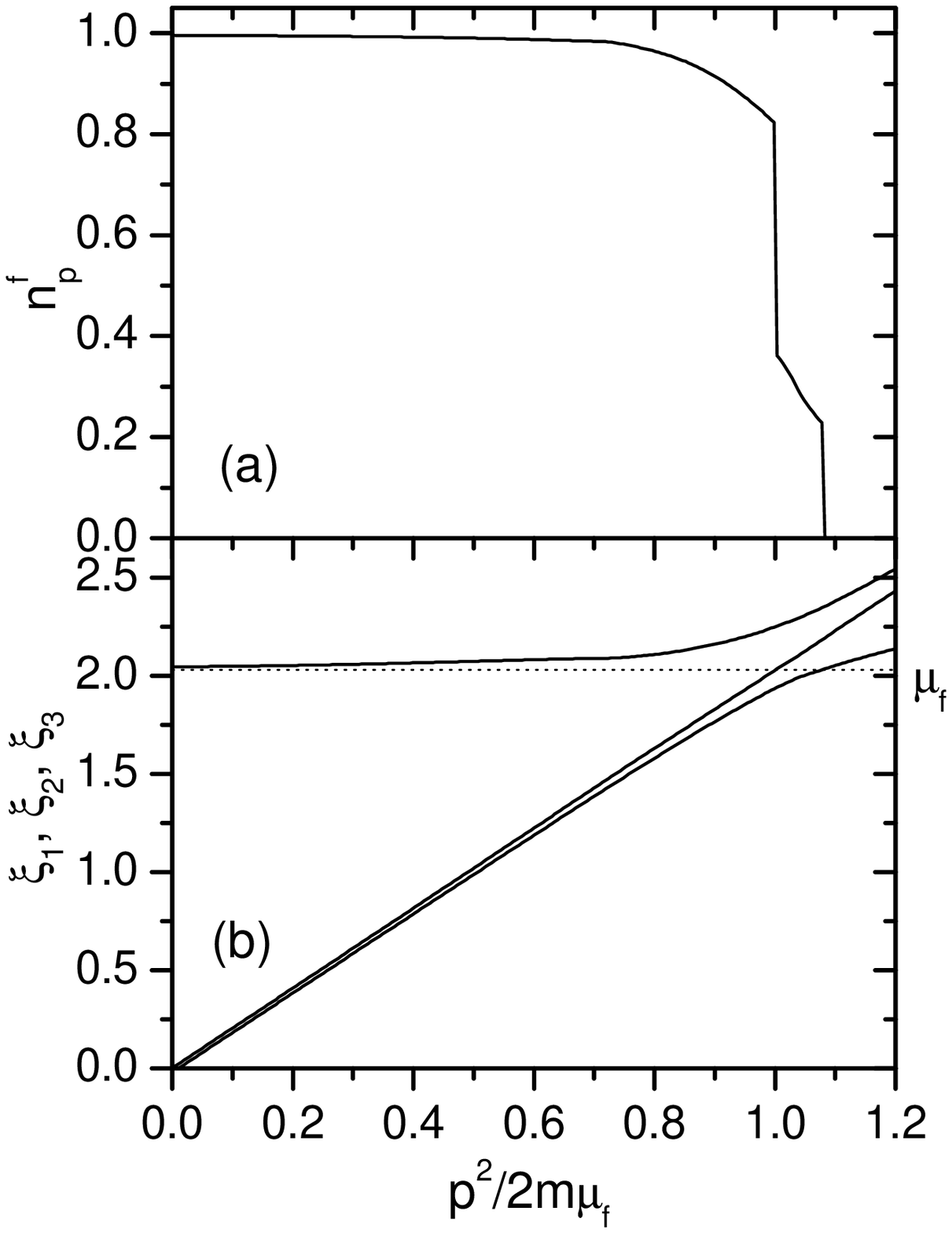}
\caption{Same as Fig. 6 but for $\protect\lambda ^{\prime }=54.0$.}
\end{figure}
With the growth of $\lambda $ the low
lying BF roots become more and more collective and when $\lambda ^{\prime }$
approaches the value equal to $\sim 54.0$ the BF roots corresponding to
total momenta $K\sim k_{F}$ become enough collective to change the Fermion
distribution strongly. In Figers 7a we thus show $n_p^f$ for $\lambda ^{\prime } = 54.0$
and the corresponding dispersion is displayed in Fig 7b.
After the level crossing, as already discussed above,
(9) and (10) exchange their properties and therefore
for $p^{2}/2m\mu _{F}\gtrsim 1$ the lowest
root becomes the collective BF pair and the highest one
becomes degenerate with the free solution.

\begin{figure}[tph]
\includegraphics[scale=0.5,angle=0]{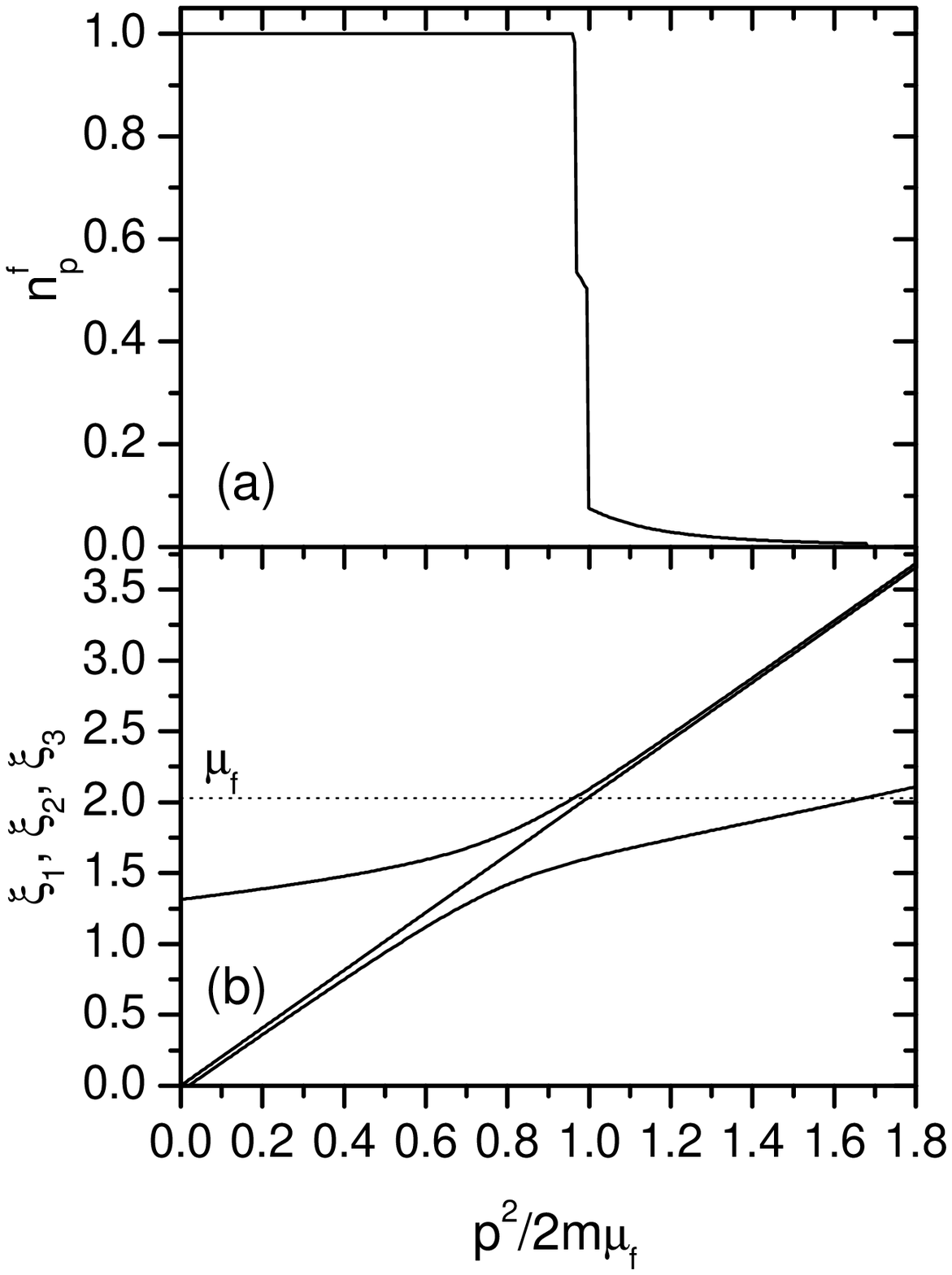}
\caption{Same as Fig. 6 but for  $\protect\lambda ^{\prime }=56.0$.}
\end{figure}
At the interaction $\lambda ^{\prime }  \gtrsim 54.0$ level crossing takes
place below the Fermi momentum.
Such a case  is displayed in Figs. 8a and 8b for $\lambda ^{\prime }= 56.0$.
We can see one small step at the momenta slightly below $k_F$,
which will disappear with
an increase of the interaction, and a rather long tail in the Fermi distribution which
corresponds to the strong collective BF pairs.
With further increasing of the interaction this tail goes to infinity whereas
the energy of the collective BF pair goes to zero.
After that the BF pair converts into the molecular state and
it is necessary to apply another kind of theory.

We
prefer not to increase $\lambda ^{\prime }$ further because the redistribution
of $n_p^f$ strongly varies with $\lambda ^{\prime }$ and
for strong $\lambda ^{\prime }$
the one pair approximation becomes invalid. On the other hand the stronger
values of $\lambda ^{\prime }$ employed in Figs. 4 and 5 have just been chosen for
illustration purposes and for a qualitative discussion of the roots
of $\mathop{\rm Re}J_{0}=-\lambda ^{-1}$
this seems quite appropriate.

We therefore very nicely see that with the parameters
chosen we have around $\mu _{F}$ a mixture of the gas of the old free
fermions and the new composite ones formed out of a boson and a fermion. The
interaction was chosen sufficiently weak so that the one pair description is
approximately valid and yet sufficiently strong so that the coexistence of
the two Fermi gases can clearly be seen.

In principle with (1), (12) and (15) one can also calculate the new boson
occupation numbers, i.e. of those bosons which, due to the B-F correlations,
are scattered out of the condensate. However, with our choice of parameters,
the influence of the bosons on the fermions remains modest and the one pair
approximation is justified. On the other hand the action of the fermions on
the bosons even for relatively small interactions is enormous because of the
much greater number of fermions than bosons invalidating the one pair
approximation. One therefore would have to iterate the calculation for the
occupation numbers,
that is we insert the new occupation numbers into (1), solve it, calculate
new occupation numbers and so on up to convergence. At the same time we will
have to reajust the chemical potentials for bosons and fermions to preserve
number of particles. This shall be done in future work and we do not show the
Boson occupation here.

The reader may have noticed that our approach is equivalent to
a particle-particle
RPA formulation \cite{cross1} in the BF channel. Since our BF pairs are
discrete states an eigenvalue variant of the present approach can give
useful additional information. We shortly present this in the Appendix.

In conclusion we considered Boson-Fermion propagation in a B-F environment
and found that the original free gas converts for sufficiently strong
attractive interaction into a completely new state of matter as a new Fermi
gas of B-F Cooper pairs with a strongly modified Fermi surface. The most
interesting and novel feature concerns the fact that, due to the Pauli exclusion
principle, this transition can occur for
interaction strength \textit{insufficient} to form bound B-F molecules in
free space. In other words the B-F pairs are at \textit{positive}
energies much analogous to Cooper pairs in a pure Fermi gas. On the other
hand the collective B-F pairs are still Fermions building a new Fermi gas of
composites and a new Fermi surface.
Whether this transition has anything to do with
the recently discovered collapse of a B-F mixture \cite{coll} remains to be
seen.

The only system parameter we varied in our work is the strength of the
interaction. Of course, a variety of other parameters could be changed: the
densities of bosons and fermions can be varied in strong proportions, their
masses could be strongly different, we worked strictly at zero temperature
only, we consider a homogeneous system and not the geometry of traps, etc.
Such investigations shall be performed in the future. It also
should again be mentioned that in this pioneering work we considered only a
very idealized situation, disregarding any interaction between like
particles, i.e. between bosons or between fermions. We suspect that as long
as the interactions between like particles are repulsive nothing qualitative
will change:
the Fermi surface will become slightly rounded and some depletion of the
condensed bosons will occur. The constellation of moderate repulsion
between particles of the same kind and attraction between different kinds
is not unrealistic \cite{hof,mil}.

In this respect we also mention that an attractive B-F interaction
can induce, via e.g. second order processes, an effective attraction
between Fermions \cite{ff}. In this paper we consider a weak coupling
scenario where
the interaction is weaker as the one needed to form a B-F molecule.
Therefore
we suppose that induced F-F attraction is weak and in any case weaker
than the direct F-F repulsion which we implicitly can assume here.
In any case, in this work we are only treating a one component
Fermi gas where s-wave scattering is suppressed, unless the force
is finite range. Then only dipole or higher odd multipole interactions
could lead to F-F attraction. On the other hand, did one consider
Fermions with spin together with bosons, F-F attraction is possible more
easily and in that case our scenario still may change strongly.
Indeed it is conceivable
that in this case standard purely fermionic Cooper pairs form new Cooper
pairs of triples in pairing up with bosons. A strong enhancement of our
present effect could occur since now 'bosons' (the F-F Cooper pairs) pair
with bosons (atoms). What exactly will happen under these conditions is
unknown at this point. Of course with attraction among Fermions, it is also
conceivable that two B-F Cooper pairs form a quartet. Those quartets would
be different from the purely fermionic quartets which may be possible when
four different species of Fermions are trapped in a pure fermi gas, as
recently discussed in the nuclear physics context ($\alpha $- particles \cite%
{ropke,alfa}). One sees that a great variety of quantum condensation
phenomena may still be explored with ultra-cold atomic gases consisting out
of Bosons and Fermions.

Note added in proof: when this work was completed we learned about the
related work in \cite{kagan} where, however, the BF pairs are treated in the
molecular state and not as collective B-F pairs embedded as a sharp
states in the continuum as in this work.

\begin{acknowledgments}
Interest in this work together with discussions by G.F. Bertsch is gratefully
acknowledged.

One of the authors (A. S.) gratefully acknowledges a post-doctoral
fellowship from the French Ministry of Research.
\end{acknowledgments}

\begin{center}
\vspace{0.8cm}
\textbf{APPENDIX}
\end{center}

A further more formal but eventually useful aspect of our theory is to
relate it to the language of a particle-particle RPA \cite{cross1} in the BF
channel. We only consider this here at $T=0$ and present the extension to
$T\neq0$ in a future publication.
To this purpose we write the following RPA excitation operator
\begin{equation}
Q_{l}^{+}=\sum_{\mathbf{pq}}\frac{X_{\mathbf{pq}}^{l}c_{\mathbf{p}}^{+}b_{\mathbf{q%
}}^{+}}{\sqrt{%
1+\delta _{\mathbf{q},0}n_{0}}}
+\sum_{\mathbf{h}}\frac{Y_{\mathbf{h}}^{l}c_{\mathbf{h}}^{+}b_{0}^{+}}{\sqrt{n_{0}}}
\end{equation}%
where $p(h)$ is a fermion momentum above (below) the Fermi sea (a 'particle'
('hole') state) and $q$ is a boson momentum which can take on all values.
With the definition of an excited state of the $N+2$ particle system (the
addition mode)
\begin{equation}
|l>=Q_{l}^{+}|>
\end{equation}%
we arrive with the usual condition
\begin{equation}
Q_{l}|>=0
\end{equation}%
at the following secular equation \cite{cross1}%
\begin{equation}
<|\left\{ \delta Q,\left[ H,Q_{l}^{+}\right] \right\} |>=E_{l}<|\left\{
\delta Q,Q_{l}^{+}\right\} |>
\end{equation}%
where $\delta Q$ is a variation with respect to either $X$ or $Y$. The usual
linearisation of (25) consists in evaluating the expectation values with the
uncorrelated ground state that is a product of a Slater determinant and\ an
ideal bose condensate. Defining the Hamiltonian as%
\begin{equation}
H=\sum_{\mathbf{q}}\left( \varepsilon _{\mathbf{q}}^{b}b_{\mathbf{q}}^{+}b_{%
\mathbf{q}}+\varepsilon _{\mathbf{q}}^{f}c_{\mathbf{q}}^{+}c_{\mathbf{q}%
}\right) +\sum_{q_{1}q_{2}q_{3}q_{4}}V_{\mathbf{q}_{1}\mathbf{q}_{2},\mathbf{%
q}_{3}\mathbf{q}_{4}}b_{\mathbf{q}_{1}}^{+}c_{\mathbf{q}_{2}}^{+}c_{\mathbf{q%
}_{3}}b_{\mathbf{q}_{4}}
\end{equation}%
where%
\[
\varepsilon _{\mathbf{k}}^{f,b}=e_{\mathbf{k}}^{f,b}-\mu _{f,b}
\]%
eq. (25) then reads as
\begin{equation}
\left(
\begin{array}{cc}
A_{\mathbf{p}^{\prime }\mathbf{q}^{\prime },\mathbf{pq}} & D_{\mathbf{p}%
^{\prime }\mathbf{q}^{\prime },\mathbf{h}0} \\
B_{\mathbf{h}^{\prime }0,\mathbf{pq}} & C_{\mathbf{h}^{\prime }0,\mathbf{h}0}%
\end{array}%
\right) \left(
\begin{array}{c}
X_{\mathbf{pq}}^{l} \\
Y_{\mathbf{h}}^{l}%
\end{array}%
\right) =E_{l}\left(
\begin{array}{c}
X_{\mathbf{p}^{\prime }\mathbf{q}^{\prime }}^{l} \\
Y_{\mathbf{h}^{\prime }}^{l}%
\end{array}%
\right)
\end{equation}%
where%
\begin{eqnarray}
A_{\mathbf{p}^{\prime }\mathbf{q}^{\prime },\mathbf{pq}} &=&\delta _{\mathbf{%
pp}^{\prime }}\delta _{\mathbf{qq}^{\prime }}\left( \varepsilon _{\mathbf{p}%
}^{f}+\varepsilon _{\mathbf{q}}^{b}\right)  \nonumber \\
&&+\sqrt{1+\delta _{\mathbf{q}^{\prime },0}n_{0}}V_{\mathbf{p}^{\prime }%
\mathbf{q}^{\prime },\mathbf{pq}}\sqrt{1+\delta _{\mathbf{q},0}n_{0}}
\nonumber \\
B_{\mathbf{h}^{\prime }0,\mathbf{pq}} &=&V_{\mathbf{h}^{\prime }0,\mathbf{pq}%
}\sqrt{n_{0}}\sqrt{1+\delta _{\mathbf{q},0}n_{0}}  \nonumber \\
D_{\mathbf{h}^{\prime }0,\mathbf{pq}} &=&V_{\mathbf{p}^{\prime }\mathbf{q}%
^{\prime },0\mathbf{h}}\sqrt{n_{0}}\sqrt{1+\delta _{\mathbf{q}^{\prime
},0}n_{0}}  \nonumber \\
C_{\mathbf{h}^{\prime }0,\mathbf{h}0} &=&\delta _{\mathbf{hh}^{\prime
}}\left( \varepsilon _{\mathbf{h}}^{f}+\varepsilon _{0}^{b}\right) +V_{%
\mathbf{h}^{\prime }0,\mathbf{h}0}n_{0}
\end{eqnarray}%
We immediately see that with $V_{\mathbf{p}^{\prime }\mathbf{q}^{\prime },%
\mathbf{pq}}=-\lambda v\left( \mathbf{p}^{\prime }-%
\mathbf{q}^{\prime }\right) v\left( \mathbf{p}-\mathbf{q}\right) \delta _{\mathbf{p}+%
\mathbf{q},\mathbf{p}^{\prime }+\mathbf{q}^{\prime }}$ the eigen values are
given as before by $\mathop{\rm Re}J_{0}=-\lambda ^{-1}$ and therefore the
RPA description is completely equivalent to the Green's function approach we
used at the beginning.

The interesting aspect of this formulation is that eq. (27) contains a
quasifermion approximation, i.e. the BF pairs in (22) are treated as ideal
Fermions%
\begin{eqnarray}
F_{\mathbf{pq}}^{+} &=&\frac{c_{\mathbf{p}}^{+}b_{\mathbf{q}}^{+}}{\sqrt{%
1+\delta _{\mathbf{q},0}n_{0}}}  \nonumber \\
F_{\mathbf{h}}^{+} &=&\frac{c_{\mathbf{h}}^{+}b_{0}^{+}}{\sqrt{n_{0}}}
\end{eqnarray}%
with
\begin{eqnarray*}
\left\{ F_{\mathbf{p}^{\prime }\mathbf{q}^{\prime }},F_{\mathbf{pq}%
}^{+}\right\} &=&\delta _{\mathbf{pp}^{\prime }}\delta _{\mathbf{qq}^{\prime
}} \\
\left\{ F_{\mathbf{h}^{\prime }},F_{\mathbf{h}}^{+}\right\} &=&\delta _{%
\mathbf{hh}^{\prime }}
\end{eqnarray*}%
This is quite in analogy to the standard RPA for a pure fermi system where a
fermion pair $c_{p}^{+}c_{p^{\prime }}^{+}$ is treated as a quasiboson \cite%
{cross1}. This quasifermion approximation contained in (27) allows to write
down the new approximate groundstate. It is given by a Slater determinant of
the new BF pairs%
\begin{equation}
|>\sim \exp \left[ \sum_{\mathbf{p}^{\prime }\mathbf{q}^{\prime }\mathbf{h}%
^{\prime }}Z_{\mathbf{p}^{\prime }\mathbf{q}^{\prime }\mathbf{h}^{\prime
}}F_{\mathbf{p}^{\prime }\mathbf{q}^{\prime }}^{+}F_{\mathbf{h}^{\prime }}%
\right] |uc>
\end{equation}%
where $|uc>$ is the 'uncorrelated' vacuum. From the condition (24) we also
can find the system of equations which defines the coefficients $%
Z_{p^{\prime }q^{\prime }h^{\prime }}$:%
\begin{equation}
Y_{\mathbf{h}}^{l}+\sum_{\mathbf{pq}}X_{\mathbf{pq}}^{l}Z_{\mathbf{pqh}%
^{\prime }}=0
\end{equation}

The initial Hamiltonian (25) in the basis of the new quasifermions (22) will be
the following%
\begin{equation}
H=\sum_{l\gtrless 0}E_{l}Q_{l}^{+}Q_{l}
\end{equation}%
where $l<0$ corresponds to the negative roots of (27) or (which is the same)
to the roots of the RPA for the $N-2$ system (the removal mode). And finally
we can define the following correlation energy%
\begin{equation}
E_{corr}=<H>-<uc|H|uc>=\sum_{l<0}E_{l}-TrC
\end{equation}%
which may be evaluated for realistic systems in the future.
This correlation energy calculated in B-F pp-RPA is exactly the analog
to the correalation energy calculated for an electron gas in ph-RPA via
the summation of ring diagrams \cite{fett}. We therefore see that
our B-F calculation also leads to an improved equation of state.

\end{document}